\documentclass[authoryear, letterpaper, 3p, 11pt]{elsarticle}
\usepackage{epsfig}
\usepackage{amsmath, amssymb}
\usepackage{fancyhdr}
\usepackage[normalem]{ulem}
\usepackage{color}
\usepackage{hanging}
\usepackage{hyperref}
\hypersetup{
  colorlinks=true,
}
\usepackage{url}
\usepackage{float, placeins}
\usepackage{mwe}
\usepackage[caption=false]{subfig}
\usepackage{morefloats}
\usepackage{empheq}
\usepackage{bm}
\usepackage{booktabs} 
\usepackage{multirow}
\usepackage{makecell}

\newcommand\Mpl{M_{\textrm{pl}}}

\usepackage{etoolbox}
\patchcmd{\MaketitleBox}{\footnotesize\itshape\elsaddress\par\vskip36pt}{\footnotesize\itshape\elsaddress\par\parbox[b][36pt]{\linewidth}{\vfill\hfill\textnormal{(Dated: November 12, 2019)}\hfill\null\vfill}}{}{}%
\patchcmd{\pprintMaketitle}{\footnotesize\itshape\elsaddress\par\vskip36pt}{\footnotesize\itshape\elsaddress\par\parbox[b][36pt]{\linewidth}{\vfill\hfill\textnormal{(Dated: November 12, 2019)}\hfill\null\vfill}}{}{}%
\patchcmd{\emailauthor}{(#2)}{}{}{}

    \makeatletter
    \def\ps@pprintTitle{%
       \let\@oddhead\@empty
       \let\@evenhead\@empty
       \def\@oddfoot{\reset@font\hfil\thepage\hfil}
       \let\@evenfoot\@oddfoot
    }
    \makeatother

\begin{document}
\hypersetup{linkcolor= blue , citecolor= magenta, urlcolor = magenta}

\title{Effective field theories as a novel probe of fine-tuning of cosmic inflation}

\author{Feraz Azhar}
\ead{fazhar@nd.edu}
\address{Department of Philosophy, University of Notre Dame, Notre Dame, IN 46556, USA\\\vspace{-0.4cm}}

\begin{abstract}
The leading account of several salient observable features of our universe today is provided by the theory of cosmic inflation. But an important and thus far intractable question is whether inflation is generic, or whether it is finely tuned---requiring very precisely specified initial conditions. In this paper I argue that a recent, model-independent characterization of inflation, known as the `effective field theory (EFT) of inflation', promises to address this question in a thoroughly modern and significantly more comprehensive way than in the existing literature. 

To motivate and provide context for this claim, I distill three core problems with the theory of inflation, which I dub the {\it permissiveness problem}, the {\it initial conditions problem}, and the {\it multiverse problem}. I argue that the initial conditions problem lies within the scope of EFTs of inflation as they are currently conceived, whereas the other two problems remain largely intractable: their solution must await a more complete description of the very early universe. I highlight recent work that addresses the initial conditions problem within the context of a dynamical systems analysis of a specific (state-of-the-art) EFT of inflation, and conclude with a roadmap for how such work might be extended to realize the promise claimed above. 
\end{abstract}

\maketitle

\tableofcontents

\section{Introduction}\label{SEC:Intro}

The most influential idea about the dynamics of the very early universe is described by the theory of cosmic inflation~\citep{guth_81}. [See also:~\citet{brout+al_78, starobinsky_80, kazanas_80, sato_81, linde_82, albrecht+steinhardt_82, linde_83}.] ``Very early'' here means before about $10^{-10}$ seconds after the putative big-bang singularity---a time that (arguably) constitutes the boundary between known and speculative physics.\footnote{Note that I will refer, in this paper, to times after some initial `big-bang singularity'. Such descriptions are, of course, imprecise. It is thought that our description of spacetime, according to general relativity, breaks down at some point in the past so as to render such times (and perhaps even the concept of a singularity) ill-defined. But I will set aside such issues for the sake of keeping the discussion self-contained.} This theory claims that for some period of time, ending at roughly $10^{-34}$ seconds after the big-bang singularity, the universe underwent accelerating expansion in which the size of the universe increased by a factor of (at least) about $e^{60}\sim10^{26}$. Preeminent among the empirical successes of inflation is its ability to provide a \emph{dynamical} mechanism that can account for small anisotropies in the cosmic microwave background  (CMB): more specifically and most prominently, it can account for angular dependences of temperature fluctuations in the CMB. Such successes have been instrumental in establishing inflation as the most popular account of very early universe cosmology (I will highlight further successes below). But, as one distinguished commentator says:
\begin{quote}
``\dots the details of inflation are unknown, and the whole idea of inflation remains a speculation, though one that is increasingly plausible''~\citep[p.~202]{weinberg_08a}.
\end{quote}
And so the question arises, how plausible is inflation? Given that inflation probes energy scales that lie beyond established physics, one way to phrase this question is to ask how probable it is for a suitable inflationary period to arise from initial conditions that we believe were possible?

Such problems of {\it fine-tuning} (as they will indeed come to be characterized in this paper) provide a rich set of open conceptual (and technical) questions that have attracted a significant amount of attention. This is partly because the fine-tuning of {\it our existence}, as encoded in our current best (effective) physical theories, is a striking putative fact. Underlying such a claim are two important foundational questions about fine-tuning that will be relevant for this paper: (i) precisely what do we mean (especially quantitatively) by fine-tuning and (ii) how might we deal with theories that predict phenomena that are finely tuned?

About (i), broadly speaking, one can identify the fine-tuning of some  salient phenomenon $\mathcal{F}$, that arises in the context of some theory $\mathcal{T}$, by the condition: were circumstances in theory $\mathcal{T}$ a little different, $\mathcal{F}$ would change significantly (for example, it would not arise at all). Of course, there is much in this characterization of fine-tuning that needs to be identified and/or made precise, for example, ``circumstances'', ``a little different'', and ``salient phenomenon $\mathcal{F}$'': in general, this can be a difficult task [see~\citet{azhar+loeb_18, azhar+loeb_19} for further context and discussion]. In this paper I will largely interpret ``circumstances'' to refer to initial conditions for dynamical variables of a theory. A colloquial understanding of ``a little different'' will mostly suffice in these introductory remarks, but I will go on to endorse a more measure-theoretic (or, with appropriate modifications, probabilistic) interpretation of this phrase---so that fine-tuning of $\mathcal{F}$ occurs when the measure over initial conditions that give rise to $\mathcal{F}$ are, in some sense, small (relative to, for example, the measure over the space of all initial conditions). A theory containing finely tuned phenomena will be referred to as a `finely tuned theory'. 

About (ii), one response that I generally endorse, is that finely tuned theories seem to call for a replacement, namely, a less-finely tuned theory, where salient phenomena do not change significantly under small changes in circumstances. Indeed, cosmic inflation provides an example of a putatively less-finely tuned theory supplanting a more-finely tuned theory, since inflation is thought to solve alleged fine-tuning problems confronting the big-bang model (BBM) of cosmology.\footnote{I refer here to the model that was well-established by the late 1970s, which traces the history of the universe back to an initial singularity without invoking an early inflationary period.} [See~\citet{mccoy_15} for another perspective on this claim.] Two such problems (that I will describe in a little more detail in the following section) are the horizon problem [see~\citet{misner_69}] and the flatness problem [see, for example,~\citet{dicke+peebles_79}]. These problems are thought to be explanatory deficiencies in the BBM's account of certain observable features of our universe today. At the heart of these explanatory deficiencies lies the contention that in the BBM, initial conditions need to be very precisely specified in order to account for these observable features. And, to be clear, it is the reliance on fine-tuning that is seen as objectionable.

Now, the question of the degree of fine-tuning of a suitable inflationary period remains an open question that has recently generated controversy---see, especially, the debate between skeptics of inflation, for example,~\citet{ijjas+al_13, ijjas+al_14} and those who remain unperturbed by such skepticism, for example,~\citet{guth+al_14} and~\citet{linde_15}. This paper clarifies issues raised in such debates and introduces a novel program that aims to resolve the controversy that surrounds one of three issues I identify, namely, the question of fine-tuning of a suitable inflationary period. 

More specifically, I will describe and endorse a new method for computing the degree of fine-tuning of cosmic inflation: a method that employs effective field theories (EFTs) to characterize inflation. I argue that this method promises to address the question of the degree of fine-tuning of inflation in a significantly more comprehensive way than in the existing literature. In order to set the stage for this argument, I review, in Sec.~\ref{SEC:virtues}, putative virtues of cosmic inflation, including aspects of the solutions provided by inflation to alleged fine-tuning problems of the BBM. I distill, in Sec.~\ref{SEC:InflationProblems}, three issues with the theory that have come to light in the recent literature, especially as a result of the debate between~\citet{ijjas+al_13, ijjas+al_14} and~\citet{guth+al_14}. I dub these problems the {\it permissiveness problem}, the {\it initial conditions problem}, and the {\it multiverse problem}. I contend that the initial conditions problem lies within the scope of EFTs of inflation, whereas the other two problems remain largely intractable---their resolution must await a more complete description of the very early universe. In Sec.~\ref{SEC:EFTI}, I provide a brief introduction to EFTs of inflation and discuss, in Sec.~\ref{SEC:EFTICP}, how they can be interpreted to address the initial conditions problem. In Sec.~\ref{SEC:InfTraj}, I highlight recent work by~\citet{azhar+kaiser_18} that provides a means to think about the onset of inflationary dynamics within the context of a (state-of-the-art) EFT of inflation. In Sec.~\ref{SEC:Road}, I outline a roadmap for how such work may be extended to realize the promise I ascribe to EFTs of inflation. Concluding remarks follow in Sec.~\ref{SEC:Conclusion}.

\section{Cosmic inflation: Claimed virtues and present shortcomings}\label{SEC:OLD}

By the late 1970s it had been recognized by some that there was an explanatory crisis in our understanding of the very early universe. The BBM of cosmology appeared to be extremely finely tuned. That is, initial conditions needed to be very precisely specified in order to give rise to the homogeneous, isotropic, and spatially flat universe we observe today. Such initial conditions consisted of (i) high degrees of initial homogeneity among patches of the universe that were, according to the BBM, causally disconnected, and (ii) a universe that was initially very close to being spatially flat. The fine-tuning inherent in (ii) arises because spatial flatness corresponds to a very special state: if the universe is initially spatially flat then it remains so, but if it begins in a spatially non-flat state (even slightly away from spatial flatness) then it rapidly diverges from flatness. The universe we observe today is thought to be spatially flat to within $0.5\%$. For this to now obtain, according to the BBM, the universe needs to be extraordinarily spatially flat early on: at roughly one second after the putative big-bang singularity, it must be flat to within $10^{-14}\%$.\footnote{In a little more detail: spatial flatness refers to a lack of curvature of the three-dimensional spatial slices that correspond to `instants' in time. It can be measured by the dimensionless ratio, $\Omega_{\textrm{tot}}\equiv{\rho_{\textrm{tot}}}/{\rho_{\textrm{c}}}$, of the total mass density in the universe, $\rho_{\textrm{tot}}$, to the critical density, $\rho_{\textrm{c}}\equiv 3H^2/(8\pi G)$ ($H$ is a measure of the expansion rate of the universe known as the Hubble parameter and $G$ is Newton's gravitational constant). Spatial flatness corresponds to $\Omega_{\textrm{tot}}=1$ (wherein the parameter in Einstein's field equations that expresses the curvature of the above-mentioned three-dimensional spatial slices, vanishes). Its present-day value has been measured to be very close to unity: $\Omega_{\textrm{tot}}=0.999 \pm 0.002$~\citep{planck_18_CP}. To obtain such a value, $\Omega_{\textrm{tot}}$ needs to satisfy---at the time of big-bang nucleosynthesis (roughly one second after the big-bang singularity)---$\Omega_{\textrm{tot}}\lesssim 1\pm O (10^{-16})$ [based on a previous estimate in~\citet{baumann_09}].}
 
\subsection{Claimed virtues}\label{SEC:virtues}

In 1980, Alan Guth introduced an idea that promised to ameliorate the fine-tuning inherent in (i) and (ii) above, through a process in which the universe underwent accelerated expansion for a very short period of time, wherein space was stretched by a factor of at least about $e^{60} \sim 10^{26}$~\citep{guth_81}. [Indeed, the title of Guth's original paper explicitly refers to the problems in (i) and (ii).] The manner in which the expansion solves these problems is well known [see, for example,~\citet{kolb+turner_90} and~\citet{linde_90}], and the solutions have also attracted philosophical attention \citep{earman_95, earman+mosterin_99, smeenk_13, butterfield_14, mccoy_15, azhar+loeb_19}. 

If all that the theory of inflation provided was a solution to fine-tuning problems with the BBM, it would probably not have the following among cosmologists that it does today. Arguably the most impressive success of the theory is its ability to provide a mechanism for understanding the origin of subtle features of the CMB---more concretely, it provides a mechanism that accounts for anisotropies in the temperature of the CMB. These anisotropies reflect inhomogeneities in the density of the universe at the time at which photons in the CMB last interacted with matter. Such inhomogeneities are thought to seed the evolution of the large-scale structures we observe today. Inflation proposes that these inhomogeneities ultimately come from small perturbations generated toward the end of inflation, arising from quantum fluctuations. [See~\citet{guth_13}, for a review.] It is a remarkable story, spanning a complex sequence of cosmological events and, if true, provides a foundation for a dynamical account of the origin of clusters of galaxies, galaxies, solar systems, stars, planets, and carbon-based life. Recent results from the {\it Planck} Collaboration provide (continued) striking confirmation of particular models that instantiate the theory of inflation in the simplest general-relativistic settings~\citep{Planck_16b_ConstraintsOnInflation, Planck_18_ConstraintsOnInflation}. 

Thus, the theory of inflation has the following two claimed virtues, which I denote with mnemonic labels since they will recur below.
\begin{quote}
(Tuning): The theory of inflation ameliorates fine-tuning problems with the BBM of cosmology, providing a dynamical mechanism to account for the high degree of homogeneity and the high degree of spatial flatness of the observable universe today.\footnote{There are other such problems that inflation provides a solution for, such as the monopole problem, but I set this aside since (Tuning), as stated, sufficiently captures features of inflation I wish to highlight.} 
\end{quote}
\begin{quote}
(Structure): The theory of inflation provides a dynamical mechanism for the origin of density perturbations in the very early universe (namely, small variations in energy density from one spatial region to another)---perturbations that are thought to grow into the large-scale structures we observe today.\footnote{Note that for the BBM of cosmology, initial conditions that could account for CMB anisotropies, mentioned above, would also need to be put in by hand---and so (Structure) can be connected to (Tuning). See~\citet{smeenk_18} for a discussion of related issues including, in particular, the role of inflation in generating initial density perturbations that can account for CMB anisotropies, as well as large-scale structure.
}
\end{quote}

\subsection{Present shortcomings}\label{SEC:InflationProblems}

Despite the evidence that confirms certain models of inflation, the theory remains speculative---and there is a relatively small band of physicists (and philosophers)  who remain skeptical of the theory. [See, for example:~\citet{earman+mosterin_99, hollands+wald_02, ellis_07, ijjas+al_13, ijjas+al_14}.] I propose to characterize such skepticism via the identification of three main problems.\footnote{Note that there are other possible problems with inflation that have received varying amounts of attention, which are not quite covered by the tripartite classification that I will develop below. Two such problems, that indeed lie outside the scope of the issues addressed in this paper, are (i) the transplanckian problem [which could be interpreted as a problem about initial conditions---but I will resist including this problem in that category (in Sec.~\ref{SEC:ICP}) for I have a more modest aim for the scope of that category] and (ii) issues to do with interpretative aspects of quantum mechanics and their impact on how we understand (Structure). See~\citet{chowdhury+al_19} (and references therein) for more  discussion of both of these issues.}

\subsubsection{The permissiveness problem}\label{SEC:PP}

The first problem, which I dub the {\it permissiveness problem}, amounts to an underdetermination of inflationary model by data. This underdetermination is daunting due to the inaccessibility of the energy scales involved and there are no agreed-upon theoretical clues about how to constrain inflationary models.\footnote{There have been various attempts to understand inflationary models in the context of theories that describe physics at higher energies, including supergravity [see, for example,~\cite{kawasaki+al_00, kallosh+linde_10, kallosh+al_13, ellis+al_13}] and string theory [see, for example,~\citet{kachru+al_03, baumann+mcallister_15}]: such theories---and the manner in which they realize inflation---remain speculative.}

The underdetermination manifests for even the simplest realizations of inflation, for which the action, $S$, can be written in the following way (where the reduced Planck mass, $\Mpl$, is given by $\Mpl \equiv 1/ \sqrt{8 \pi G} \approx 2.4 \times 10^{18}$ GeV, in units where $\hbar=c=1$):
\begin{equation}\label{EQN:ActionSFI}
S = \int d^{4}x\sqrt{-g}\left[\frac{\Mpl^{2}}{2}R-\frac{1}{2}g^{\mu\nu}\partial_{\mu}\phi\partial_{\nu}\phi -V(\phi)\right].
\end{equation}
This action contains five salient features [see, for example,~\citet[Ch.~2]{pieroni_16}]: 
\begin{itemize}
\item[(i)] a single-scalar field, $\phi(t, \vec{x})$, known as the `inflaton field', or simply the `inflaton' (represented here as a function of a cosmic time coordinate, $t$, and, implicitly, three spatial coordinates, $\vec{x}$); 
\item[(ii)] a potential term, $V(\phi)$, that plays a key role in determining dynamical properties of the inflaton; 
\item[(iii)] the use of a canonical form for kinetic degrees of freedom for the inflaton [as in the second term in brackets in Eq.~(\ref{EQN:ActionSFI})]; 
\item[(iv)] an account of gravitational degrees of freedom via the Einstein-Hilbert action [that is, the first term in brackets in Eq.~(\ref{EQN:ActionSFI})]; 
\item[(v)] minimal coupling between the inflaton and the gravitational degrees of freedom as represented in (iv) [in effect, no term in the action that conjoins functions of the Ricci scalar, $R$, and the inflaton (and/or functions of their derivatives)]. 
\end{itemize}
Inflationary models that invoke all five features will be referred to as `SSF' (for `single-scalar field') models of inflation.

Different SSF models of inflation can be constructed through different choices for the potential $V(\phi)$; it turns out that there are a plethora of such models that realize (Tuning) and (Structure). A particularly striking account of the multiplicity of these models is provided by~\citet{martin+al_14a} who analyzed 74 distinct inflaton potentials that have been proposed in the literature: again, each of them corresponding to an SSF model of inflation. In a Bayesian study comparing such models with the Planck satellite’s 2013 data about the CMB~\citep{martin+al_14b}, a total of 15 different potentials are favored by the data; this number can be reduced to 9 different potentials if a particular measure of complexity is invoked. So although this represents a significant reduction in the total number of SSF models that are consistent with data, an underdetermination problem remains. 

And it is important to note that SSF models belong to just one class of models of inflation. For instead of probing different potentials [namely, different choices for (ii) above], one could look to vary each of the other components. Such an exploration is more than just an exercise in listing possibilities generated by denying some combination of (i), (iii), (iv), and (v): such possibilities do indeed arise when one looks to embed inflationary models in (admittedly, speculative) higher-energy theories. [See, for example:~\citet{amendola_93, wands_08, defelice+tsujikawa_10, kaiser_16}.]

I'll mention two such examples that  focus on different aspects of the assumptions that underlie SSF models. First, one may look to modify the {\it matter content} of the early universe in such a way as to relax the choice in (i). That is, it is thought that around the time of inflation, there may have existed multiple scalar fields. One is thus led to models of `multi-field inflation', which can have consequences for observables today that are distinct from those derived from models involving single-scalar fields~\citep{wands_08}.\footnote{See also~\citet{senatore+zaldarriaga_12}, who describe novel features of the inferred statistical description of density perturbations generated during inflation---in particular, novel aspects of non-Gaussianities.} Second, one may look to modify {\it gravitational degrees of freedom}, that is, to modify the Einstein-Hilbert action. One class of such modifications introduces higher-order curvature invariants in the action. A well-known modification of this form is that of $f(R)$ gravity, where some (for example, nonlinear) function of the Ricci scalar appears in the action~\citep{defelice+tsujikawa_10}. Indeed one of the earliest inflationary models was of this type, based on a modification of the Einstein-Hilbert action by~\citet{starobinsky_80}, with $f(R)=R+\alpha R^2$ (where $\alpha$ is a constant). Strikingly, this model furnishes predictions for observational parameters that are in excellent agreement with the Planck data.\footnote{\label{FN:Star}In particular, this model is in agreement with data on the scalar spectral index $n_s$ and the tensor-to-scalar ratio $r$. Note also that one can perform a conformal transformation that reinterprets this modification of the Einstein-Hilbert action in terms of a scalar degree of freedom subject to a potential [which I'll denote by $V_{S}(\phi)$]. An analysis of interpretational issues that arise for such a  mapping---namely, between the `Jordan frame' and the `Einstein frame'---lies outside the scope of this paper.} And, as suggested above, these two types of modifications to SSF models of inflation don't include other possibilities that could be considered (and have indeed been pursued), such as noncanonical forms for kinetic degrees of freedom, or single- or multi-field models that are nonminimally coupled to gravitational degrees of freedom\footnote{A particularly interesting inflationary model, known as `Higgs Inflation', is of precisely this character. Here a single-scalar field is nonminimally coupled to the Ricci scalar, with a (Jordan-frame) potential that is of the familiar Higgs variety. Under a conformal transformation, one can derive a model (in the Einstein frame) with a minimally coupled single-scalar field and a potential that corresponds (in a certain limit) to $V_{S}(\phi)$ (in fn.~\ref{FN:Star}). [See, for example,~\citet{bezrukov+shaposhnikov_08}.]}. 

The upshot is that there is, at present---whether one focuses on the simplest realizations of inflation or on extensions of such realizations---a severe underdetermination problem of inflationary model by the best cosmological data that we have available. 

One of the motivations for studying EFTs of inflation is as a response to the permissiveness problem. EFTs of inflation describe an inflationary phase in a way that is common to the class of inflationary models over which they are agnostic:  as such, they may be able to ameliorate the permissiveness problem (I will touch upon this again in Sec.~\ref{SEC:EFTICP}). 

\subsubsection{The initial conditions problem}\label{SEC:ICP}

The second problem amounts to the question of whether inflation is finely tuned.\footnote{One can (but, strictly speaking, one does not have to) reframe this question as one that probes whether inflation is {\it less-}finely tuned than the BBM---thereby calling into question one aspect of (Tuning).} Indeed, the question of whether inflation requires very precisely specified initial conditions (that is, whether inflation is finely tuned with respect to initial conditions)---what I will call the {\it initial conditions problem}---is controversial [\citet{brandenberger_17} provides an overview of recent progress; see also~\citet{linde_15, linde_18}].\footnote{\citet{penrose_89} characterizes the fine-tuning problem for inflation in terms of entropy considerations.} There are two natural issues relevant to the initial conditions problem: first, an issue related to when inflation began (in a sense, the issue concerns how we specify ``initial'') and secondly, the issue of how we characterize the ``conditions'' in which we are interested. 

With regard to the first issue, consider the following two possible scenarios.
\begin{itemize}
\item[(I)] Inflation began immediately after the universe emerged from the Planck era, namely, at the Planck time of about $10^{-43}$ seconds after the big-bang singularity. In this case, `initial conditions for inflation' refers to conditions of the universe at the Planck time. 
\item[(II)] There was a delay after the universe emerged from the Planck era, before inflation began. In this case, `initial conditions for inflation' could refer to: (a) the same initial conditions as for (I), namely, conditions of the universe at the Planck time; or (b) conditions of the universe just before inflation indeed begins. 
\end{itemize}
In the simplest (that is, SSF) realizations of inflation (as described above)---where the potential $V(\phi)$ is assumed to govern the potential-energy density of the universe immediately after the Planck era---general arguments show that scenario (I) renders inflation less-finely tuned than in scenario II(a). In particular, a smaller initial patch of homogeneity is required to initiate inflation in (I). Inflation ``protects'' patches of homogeneity against surrounding inhomogeneities, which can evolve to overwhelm the homogeneous patch and prevent it from inflating [see, for example,~\citet[Sec.~3.4.4]{liddle+lyth_00}]. Option II(b) can naturally be interpreted as treating inflation as an EFT (as we will in Sec.~\ref{SEC:PromEFT}), in which the details of the theory are somewhat independent of higher-energy (for example, Planck-energy-scale) completions of the theory~\citep{weinberg_08b, cheung+al_08a}.

With regard to the second issue, for SSF models of inflation (setting aside where one stands on claims related to the first issue as described in the previous paragraph), there are two main classes of ``conditions'' that have been probed in the investigation of how inflation depends on initial conditions.\footnote{I set aside related issues concerning the (classical) stability of de Sitter space (to perturbations)---as addressed in cosmic `no-hair' conjectures---see, for example:~\citet{gibbons+hawking_77},~\citet{hawking+moss_82},~\citet{wald_83},~\citet{barrow_83},~\citet{boucher+gibbons_83},~\citet{starobinsky_83},~\citet{jensen+steinschabes_87},~\citet{barrow_87},~\citet{kleban+senatore_16}; for a concise conceptual overview, see~\citet[Sec.~5.3.4]{barrow_17}.}
\begin{itemize}
\item[(i)] The first approach investigates how precisely initial conditions need to be set to yield sufficient amounts of inflation starting from cosmological settings that are {\it homogeneous and isotropic} [namely, Fredmann-Lema\^{i}tre-Robertson-Walker (FLRW) spacetimes]. In this case, one typically considers the standard FLRW metric in the context of a homogeneous scalar field: $\phi(t,\vec{x})=\phi(t)$. A further simplification can be made where one assumes that spatial sections (as encoded in the FLRW metric) are initially flat. This approach, in effect, sets aside some of the motivations for inflation described above [for example, in (Tuning)].
\item[(ii)] The second approach---which addresses a larger challenge---investigates how precisely initial conditions need to be to set to yield sufficient amounts of inflation starting from cosmological settings that are inhomogeneous and/or anisotropic. This second line of attack has the potential to address the more conceptual question of whether inflation provides a better \emph{explanation} of observed features of our universe today (by virtue of it being less-finely tuned) than the BBM. [See~\cite{azhar+loeb_19} for a recent discussion of the relationship between fine-tuning and {\it explanatory depth}.]
\end{itemize}

The consensus amongst studies that address both (i) and (ii) appears to be that large-field inflation models---that is, those in which the excursion of the scalar field during inflation, $\delta\phi$, is greater than the Planck mass ($m_{\textrm{Pl}}$), $\delta\phi \gtrsim m_{\textrm{Pl}}$, are more robust to changes in initial conditions than are small-field inflation models (where  $\delta\phi \ll m_{\textrm{Pl}}$ during inflation). In the case of (i), for example,~\citet{remmen+carroll_14} show by explicitly constructing a measure on the space of trajectories for flat FLRW spacetimes, that if the initial energy density of the universe is the Planck energy density, then for a quadratic potential, $V(\phi)=\frac{1}{2}m^2\phi^2$ (an example of a potential consistent with large-field inflation), nearly all trajectories undergo sufficient amounts of inflation. For cosine inflation, $V(\phi)=\Lambda^4\left[1-\cos(\phi/f)\right]$ (an example of a potential consistent with small-field inflation), one finds the opposite conclusion.\footnote{In the case of cosine inflation, this conclusion is sensitive to the value of $f$. Their conclusions, as reported above, assume that relevant parameters for the two potentials are consistent with observational data (that is, data from the {\it Planck} Collaboration).} In the case of (ii) above, recent numerical work (in the context of 3+1-dimensional Einstein gravity) establishes a similar conclusion in the case where inhomogeneities are included [see:~\citet{east+al_16, clough+al_16, clough+al_18, marsh+al_18, bloomfield+al_19, aurrekoetxea+al_19}]. [See also~\citet{chowdhury+al_19}, for a discussion of the above issues.]

Note however that these results generally relate to just one way in which inflation can be realized (that is, via SSF models), and the size and nature of the inhomogeneities probed have limitations. I will argue below that there are decidedly distinct ways to supplement and extend such results so as to provide a more complete picture of the severity of the initial conditions problem for inflation.

\subsubsection{The multiverse problem}\label{SEC:MVprob}

The third problem is related to a striking prediction of inflation, if it is extended to energy scales higher than those involved in the dynamical imprinting of density perturbations in the very early universe [as mentioned in (Structure)]. Namely, it is thought that under such an extension, inflation is future-eternal and gives rise to a multiverse: a vast (typically infinite) causally disconnected set of `pocket universes' in which the parameters of the standard models of particle physics and cosmology vary from one pocket universe to the next~\citep{vilenkin_83, linde_86a, linde_86b, guth_07, guth_13}.\footnote{Here, by the `standard model of cosmology' I am referring to the $\Lambda$CDM model of cosmology---so named after the components that are thought to dominate the energy density of the universe, namely, the cosmological constant (represented by $\Lambda$) and cold dark matter (CDM).} The theories that include such extensions are known as theories of `eternal inflation'.  

A question that arises in this context is: how do we test some theory that describes a multiverse? The natural approach is to determine predictions, extracted from the theory, for what {\it we} would observe (in our pocket universe). Such a test can be operationalized through a comparison between {\it probability distributions}, as constructed from the theory, and our observations. But in constructing such  distributions one must confront a variety of challenges that span both technical and conceptual issues. One can parse these challenges into three sub-problems: the measure problem, the conditionalization problem, and the typicality problem. The conjunction of these three sub-problems will be referred to as the {\it multiverse problem}.\footnote{See~\citet{aguirre_07} and~\citet{azhar+butterfield_18} for more detailed discussion. In what follows I will summarize key aspects of this problem.}

The measure problem amounts to two issues. First, one needs to specify the sample space, that is, the collection of `outcomes'---sets of which can receive a measure. But there is no a priori best way of selecting the sample space. Secondly, even if one has specified a reasonable sample space, natural measures over appropriate sets of outcomes can be infinite, making it difficult to define probabilities (such as when probabilities are taken as ratios of measures). Regularization schemes have indeed been introduced with the goal of taming such infinities but there is no single agreed-upon scheme, and resulting probabilities indeed depend upon the scheme employed. [See, for a discussion of these as well as other challenges:~\citet{aguirre_07, aguirre+al_07, guth_07, desimone+al_08, freivogel_11, smeenk_14}.]

Now, it is likely that even with a solution to the measure problem in hand, probabilities for physical quantities taking the values that {\it we} measure will be small: for generic models of eternal inflation, much of the multiverse is not likely to resemble our pocket universe. So instead of simply rejecting some such model of eternal inflation on the grounds that it does not predict, with high probability, (our) observed values for salient physical quantities, one {conditionalizes} the relevant probability distribution with respect to some set of criteria $\mathcal{C}$ (that is, one implements a `conditionalization scheme'). Such criteria are meant to reflect our `observational situation' or perhaps `us', but it is not clear how to do this. A natural choice for $\mathcal{C}$ is that it comprises a comprehensive account of, for example, our observational situation, as in {\it all the data we have thus far accumulated}, but it is not clear how we can practically implement such a choice. Of course, different choices for $\mathcal{C}$ lead to different probabilities for physical quantities and thus to different predictions for what we should expect to observe [viz.~a version of the `problem of the reference class': see~\citet{hajek_07}]. This problem, of which criteria $\mathcal{C}$ to choose and then how to practically conditionalize the relevant probability distribution based on this choice is the conditionalization problem~\citep{aguirre+tegmark_05}. 

Finally, to elicit a precise prediction---even if one has determined a measure and a conditionalization scheme---one must confront the question of how typical we should expect to be of spacetime regions so delineated (that is, spacetime regions in which we might arise), and therefore how typical we should expect the values of our observed physical quantities to be. The assumption that we \emph{are} typical  is known as the principle of mediocrity~\citep{vilenkin_95}. But this assumption has been controversial. Precisely what assumption we should adopt is the `typicality problem'. A more complete discussion of these interesting sub-problems will take us too far afield, but for a growing literature on such issues (and especially the typicality problem), see:~\citet{weinstein_06},~\citet{hartle+srednicki_07},~\citet{garriga+vilenkin_08},~\citet{srednicki+hartle_10},~\citet{azhar_14, azhar_15, azhar_16, azhar_17},~\citet{azhar+butterfield_18}.

Of the three present shortcomings described above---namely, the permissiveness problem, the initial conditions problem, and the multiverse problem---I contend that the initial conditions problem is the most pressing. Namely, the possibility that inflation is finely tuned presents a significant problem for the theory. With regard to the permissiveness problem there is a hope (which I generally endorse) that future observational evidence and theoretical advances will discriminate between various possible realizations of inflation. Furthermore, the fact that different ways of realizing inflation can account for different (future) observational outcomes points to a robustness (rather than a failing) of the theory. [See, for example,~\citet{guth+al_17}, for thoughts along these lines.] The multiverse problem pushes back our description of the very early universe to energy scales that outstrip those of the last phase of inflation itself [the phase that has observable consequences encapsulated especially in (Structure)], and is thus arguably less pressing (for now) than the other two.

\section{The promise of effective field theories of inflation}
\label{SEC:PromEFT}

In the past decade or so, a promising body of work has been developed that probes the idea of cosmic inflation in a way that is significantly different from earlier characterizations. As I will argue below, this new perspective on inflation as an effective field theory (EFT)~\citep{cheung+al_08a, weinberg_08b} may well provide a novel way to tackle the initial conditions problem. 

Broadly speaking, EFTs of inflation probe a general characterization of inflation around a preferred energy scale, where details of a more fundamental theory that may underlie this characterization are, in effect, suppressed. This energy scale turns out to be significantly lower than the Planck energy scale [due to observational constraints, see~\citet{Planck_16b_ConstraintsOnInflation, Planck_18_ConstraintsOnInflation}]. The approach, endorsed by~\citet{guth+al_14}, and developed by~\citet{cheung+al_08a} and~\citet{weinberg_08b} connects to one of the more consequential shifts across many areas of theoretical physics over the past 45 years, namely, a shift toward description in terms of EFTs---as has occurred in high-energy physics and condensed-matter physics. Thus, a promising next step in developing an understanding of the very early universe could well be to think in similar terms, about dynamics through effective degrees of freedom that are most relevant within some energy range. Also, for very early universe cosmology, there is an additional reason to adopt an EFT-style approach, which is that information about physics at considerably higher energies would presumably be inaccessible to observation, even in principle. For if inflation did occur, then  information about how the universe behaved before the final phase of inflation (namely, the phase that we can indeed probe through our observations today), would have been stretched to length-scales that would now be much longer than those that describe the universe today. [This last point, of course, sets aside the manner in which one might construe tests of the multiverse (described in Sec.~\ref{SEC:MVprob}) as providing information about physics at such higher energies.]

The adoption of the spirit promoted in the previous paragraph comes with the added feature that EFTs of inflation [especially as pioneered by~\citet{cheung+al_08a}] provide {\it model-independent} characterizations of inflation (around the preferred energy scale). As I will argue below, this perspective on inflation does not resolve the permissiveness problem (though there is a sense in which it can address an aspect of the problem) or the multiverse problem, but it does provide a novel angle on the initial conditions problem and, moreover, promises to address this problem in a significantly more comprehensive way than in the existing literature. 

\subsection{Effective field theories of inflation: A brief introduction}\label{SEC:EFTI}

EFTs of inflation achieve model-independence by characterizing inflation through symmetries of the underlying spacetime. In the language of classical physics, inflation is described as a period of accelerating expansion of the universe---together with small fluctuations atop a homogeneous and isotropic background---where the universe is in a quasi-de Sitter state: it is not in an exact de Sitter state because the accelerating expansion must end. In this way, there is a symmetry of general relativity that must be broken, namely, time-translation invariance of the dynamics, as encoded in the relevant action. In more practical (and yet still model-independent) terms, the early universe contains a `clock' that counts down time until the end of inflation. Importantly, this clock does not have to correspond to the usual scalar field $\phi$ (as in the SSF models described in Sec.~\ref{SEC:PP}), and thus EFTs of inflation probe realizations of inflation that extend beyond the usual settings in which it has been studied. Inflation can thus be thought of  as a (particular) theory of 4-dimensional spacetime diffeomorphisms broken to time-dependent, 3-dimensional spatial diffeomorphisms. The construction of the action for such a scenario is described in, for example,~\citet{cheung+al_08a} and~\citet[Appendix B]{baumann+mcallister_15}. Only the rudiments of such a construction will be needed for our discussion, to which I now turn.  

One can write down the most general effective action respecting time-dependent spatial diffeomorphisms using a particular time-slicing of the underlying spacetime. This time-slicing is known as `unitary gauge' and corresponds to a choice of coordinates such that fluctuations in the clock at different spatial locations vanish (to first order), leaving only perturbations in the metric. The action---assuming we expand around a spatially flat FLRW metric, where the scale factor is denoted by $a(t)$---can be written in the following way:
\begin{equation}\label{EQN:ActionFO}
S = S_0 + \Delta S,
\end{equation}
where
\begin{equation}\label{EQN:S0}
S_0 =\int d^{4}x\sqrt{-g}\left[\frac{\Mpl^{2}}{2}R-L(t)-c(t) g^{00}\right].
\end{equation}
In Eq.~(\ref{EQN:ActionFO}), $\Delta S$ contains two expansions, that is, it contains (i) terms that are at least quadratic in fluctuations of the metric and (ii) terms that contain higher-order derivatives of the metric. These latter terms correspond to the types of terms that arise in more standard EFT settings and (as in those settings) are assumed to be suppressed in the low-energy effective theory. When one is interested in dynamical aspects of just the background spacetime (as indeed we will be below, and especially in Sec.~\ref{SEC:InfTraj}), one can ignore $\Delta S$. In Eq.~(\ref{EQN:S0}), the two functions, $L(t)$ and $c(t)$, are (thus far) unspecified functions of cosmic time: they can be given a natural interpretation (as I will develop shortly) and will play a crucial role in the dynamical analysis of this EFT of inflation, as described in Sec.~\ref{SEC:InfTraj}.

One can derive Einstein's equations (that is, in this specific context, Friedmann's equations), which encode the dynamics of the background spacetime, in the usual way, by varying $S_0$ with respect to $g^{\mu\nu}$. This yields
\begin{align}
H^{2}(t)&=\frac{1}{3 \Mpl^{2}}\left[c(t)+L(t)\right], \label{EQN:EE1}\\
\dot{H}(t)+H^{2}(t)&=-\frac{1}{3 \Mpl^{2}}\left[2 c(t)-L(t)\right], \label{EQN:EE2}
\end{align}
where $H(t) \equiv \dot{a}(t) / a(t)$ is the Hubble parameter (overdots denote derivatives with respect to cosmic time). We thus have expressions that relate the dynamics of the background---namely, of the scale factor $a(t)$---to the functions of cosmic time, $L(t)$ and $c(t)$, appearing in $S_0$.  We also note that solving Eqs.~(\ref{EQN:EE1}) and (\ref{EQN:EE2}) for $c(t)$ and $L(t)$ yields 
\begin{equation}\label{EQN:cLSolve}
c(t)=-\Mpl^{2}\dot{H}(t),\qquad
L(t)=\Mpl^{2}\left[3H^2(t)+\dot{H}(t)\right].
\end{equation}

To connect such a description of inflation with the simplest realizations of inflation, compare the action in Eq.~(\ref{EQN:S0}) with the action for SSF models of inflation [Eq.~(\ref{EQN:ActionSFI})] in unitary gauge. In this gauge, fluctuations in the clock (now, the inflaton) vanish (to first order), so we can set $\phi(t,\vec{x})\to\phi_{0}(t)$. Equation~(\ref{EQN:ActionSFI}) becomes
\begin{equation}\label{EQN:ActionSFIUG}
S = \int d^{4}x\sqrt{-g}\left[\frac{\Mpl^{2}}{2}R-\frac{1}{2}g^{00}\dot{\phi}_{0}^2(t)-V(\phi_{0}(t))\right].
\end{equation}
Equation~(\ref{EQN:ActionSFIUG}) has precisely the same form as Eq.~(\ref{EQN:S0}) if we identify
\begin{equation}\label{EQN:VLphidotc}
 \frac{1}{2} \dot{\phi}_0^2 (t) \leftrightarrow c (t),\qquad
V (\phi_0 (t)) \leftrightarrow L (t).
\end{equation}
Thus the action for an SSF model, in unitary gauge, corresponds to the action displayed in Eq.~(\ref{EQN:S0}), though the action in Eq.~(\ref{EQN:S0}) is not limited to the case of SSF models~\citep{cheung+al_08a}. 

It is worth reiterating this point. That is, the  EFT-of-inflation action for the background spacetime, displayed in Eq.~(\ref{EQN:S0}), contains functions of time, namely, $L(t)$ and $c(t)$, that are {\it agnostic} about the underlying physical mechanism that gives rise to these functions---it is in this sense that EFTs of inflation provide a model-independent characterization of inflation.\footnote{Note that there is a second sense of model-independence also in play, in that the full EFT expansion in Eq.~(\ref{EQN:ActionFO}) is agnostic about details of a more fundamental theory that may describe inflation at higher energy scales. But it is the sense of model-independence in the main text that will be important in what follows.}

\subsection{Effective field theories of inflation and present shortcomings of inflation}\label{SEC:EFTICP}

A question that now arises is: how does such an EFT of inflation relate to the shortcomings described in Sec.~\ref{SEC:InflationProblems}? I will provide brief responses for two of these problems---namely, the multiverse problem and the permissiveness problem---before focusing on the initial conditions problem. 

EFTs of inflation, almost by construction, do not address the multiverse problem. This is because EFTs of inflation apply at a preferred energy scale that is lower than that typically employed to probe multiverse cosmological scenarios. Also, EFTs of inflation probe small fluctuations atop an inflating background, whereas, at least in particular variants of eternal inflation (namely, slow-roll eternal inflation), much larger fluctuations need to be addressed. On both accounts---that of the preferred energy scale of the theory and the nature of the fluctuations invoked---EFTs of inflation are not well-suited to an analysis of an inflationary multiverse.

The relationship between EFTs of inflation and the permissiveness problem is more nuanced. EFTs of inflation are model independent (in the sense specified at the end of Sec.~\ref{SEC:EFTI}) and thus they aim to characterize what is common among models that are (partly) responsible for the charge that inflation is too permissive. But this provides an opportunity in that, for example, if one can generate a prediction from an EFT of inflation that is not borne out in experiments, then one can disconfirm each model in the class of models captured by the EFT. In which case, EFTs of inflation may be able to ameliorate the permissiveness problem. See, for example,~\citet{cheung+al_08b}, who describe such a possibility in the case of EFTs of inflation that invoke a single clock.

More generally, however, a resolution of the permissiveness problem and the multiverse problem awaits a more complete theory of the very early universe. Indeed, if we were able to formulate such a theory, it may pick out a particular model by which inflation was realized. Additionally, such a theory would furnish an understanding of dynamics at higher energy scales and thus may also contain an account of whether a multiverse can arise (along with an account, for example, of its observable features). Of course, with due respect to a corresponding problem of initial conditions for this more complete theory, it would also presumably provide insight into the initial conditions problem for inflation at an energy scale that is accessible, in principle, to observations today. However, we do not have such a complete theory, and part of my point in this paper is that we do not have to wait for its development, in order to address the initial conditions problem as it pertains to an inflationary period that could have yielded the observable universe: EFTs of inflation promise to provide a rather comprehensive account of this issue.

Thus my main claim is that EFTs of inflation do provide an opportunity to address the initial conditions problem for inflation---and they do so in a way that is significantly different from all attempts to address this problem thus far. As mentioned above, EFTs of inflation focus on preferred energy scales in such a way that they answer, by design, the preliminary issue relevant to the initial conditions problem of inflation discussed in Sec.~\ref{SEC:ICP}. Namely ``initial'' in ``initial conditions'' is construed as ``around the preferred energy scale for the EFT''. Of  course, this restricts the scope of an EFT of inflation in addressing the initial conditions problem---for there remains the question of how precisely initial conditions need to be specified at the Planck energy scale, regardless of what an EFT-of-inflation analysis reveals about the nature of initial conditions at lower energy scales.

To further delineate the scope of EFTs of inflation in addressing the initial conditions problem, I highlight two aspects of this problem that were left implicit in the discussion in Sec.~\ref{SEC:ICP}, and give them mnemonic labels as they will recur.
\begin{itemize}
\item[] (Characterize): First is the issue of how one characterizes inflation. On this score, {\it every} attempt so far to probe the initial conditions problem [save for~\citet{azhar+kaiser_18}] has assumed a specific functional form for the inflaton potential. In the scenarios summarized in Sec.~\ref{SEC:ICP}, the analytical work [by, for example,~\citet{remmen+carroll_13, remmen+carroll_14}] as well as recent numerical work [by, for example,~\citet{east+al_16, clough+al_16, clough+al_18, marsh+al_18, bloomfield+al_19, aurrekoetxea+al_19}] invokes specific potentials (admittedly, these potentials are generally taken to be representative of one of two classes of potentials---that is, large-field models or small-field models). Their conclusions are thus potential-dependent or, at best, depend on the classes into which the potentials fall. [See also~\citet{chowdhury+al_19}.]
\item[] (Measure): The second issue concerns how one would unambiguously verify the existence of a problem in the first place, that is, how one would verify the existence of fine-tuning. On this score---and in perhaps what is the least tractable aspect of the initial conditions problem (regardless of how one characterizes inflation)---work generally focuses on choosing a measure over initial conditions as they arise in some phase space for the theory. From this measure, one would like to compute, for example, the probability of sufficient amounts of inflation occurring. The problem here is that the choice of such a measure is not uniquely selected by any underlying theory. Moreover, reasonable measures over appropriate phase spaces are often infinite, making it difficult to define probabilities (as for the measure problem that arises as part of attempts to extract predictions from theories of the multiverse, as discussed in Sec.~\ref{SEC:MVprob}). [See~\citet{schiffrin+wald_12} and~\citet{smeenk_14} for further discussion.] The initial conditions problem, interpreted in a measure-theoretic context, thus includes a very thorny measure problem.\footnote{And to be clear, this measure problem is distinct from the measure problem that arises in analyses of multiverse scenarios.}
\end{itemize}

EFTs of inflation most naturally provide a novel perspective on the first of these two issues. Namely, their model-independence provides a new way of characterizing inflation. As mentioned above, the EFT of inflation developed by~\citet{cheung+al_08a} includes all models of inflation that involve a single clock that counts down time until the end of inflation. This includes the clock that is employed by the studies mentioned under (Characterize) above (that is, an SSF $\phi$), but also any other way that inflation might be realized via a single clock, including ways for which we have not yet found any sort of physical underpinning. The promise of EFTs of inflation is that one might be able to identify relevant phase spaces that are suitably agnostic about the mechanism that underlies inflation [unlike each study, save for~\citet{azhar+kaiser_18}, mentioned in (Characterize)] so that measures over such phase spaces [that is, measures that arise from a solution to the issue described in (Measure)] will be able to guide a more comprehensive assessment of the nature and severity of the initial conditions problem. 

If we are indeed able to identify suitable phase spaces that describe single-clock inflation and also construct well-motivated measures over such phase spaces (not an easy task), the conclusions we would reach about whether inflation, so characterized, is generic, would arguably be an important guide to determining whether inflation is indeed generic. In the following section, I describe and endorse an attempt to achieve rudiments of such a construction. 

\subsection{A dynamical systems analysis of an effective field theory of inflation}\label{SEC:InfTraj}

In \citet{azhar+kaiser_18} we combine recent work on EFT approaches to inflation with a dynamical-systems analysis originally formulated to characterize late-universe acceleration~\citep{frusciante+al_14}. In short, we develop tools with which to address the flow into inflationary states for the (general) single-clock descriptions of inflation detailed above---a formulation that includes, but is not limited to, SSF models of inflation. In order to develop the formalism, we restrict attention to background spacetimes that are (already) homogeneous, isotropic, and spatially flat, focusing on the dynamical flow into inflationary states for initial conditions that are not expressly geared to trigger inflation. We develop heuristic measures over such initial conditions with which we estimate the probability that inflation will begin and persist for at least 60 $e$-folds. 

In more detail then, the physical system of interest is a flat background space together with a clock that counts down time until the end of inflation, as described in Sec.~\ref{SEC:EFTI}. There are, in principle, three key variables: the scale factor $a(t)$, which characterizes the background space; and two functions of cosmic time, $c(t)$ and $L(t)$, which characterize the clock (a further interpretation of these functions of time, in the context of SSF models of inflation, appears below).  One can expressly convert the EFT-of-inflation formalism into a dynamical system by adapting techniques described by~\citet{frusciante+al_14}. In particular, one defines the following dimensionless variables (suppressing explicit time dependences from now on), which serve as dynamical variables for the resultant dynamical system:
\begin{equation}\label{EQN:IF}
x\equiv \frac{c}{3 \Mpl^{2} H^2},\qquad 
y\equiv \frac{L}{3 \Mpl^{2} H^2},\qquad
\lambda_{m}\equiv-\frac{L^{(m+1)}}{H L^{(m)}},
\end{equation}
for $m=0,1,2,\dots$. In the third expression in Eq.~(\ref{EQN:IF}), $(m)$ represents the $m$th derivative with respect to time. This expression introduces an infinite tower of dimensionless variables that encode implicit choices for the functional form of $L$, though, as I will describe below, in practice we only consider a finite number of these terms for a given phase-space analysis. 

Note that for the first Einstein equation to be satisfied [Eq.~(\ref{EQN:EE1})] the dimensionless variables $x$ and $y$ must satisfy a constraint:
\begin{equation}\label{EQN:Constraint}
1=x+y.
\end{equation}
A further feature of these dynamical variables follows from Eqs.~(\ref{EQN:EE1}) and~(\ref{EQN:cLSolve}), namely that the Hubble slow-roll parameter $\epsilon$, which delineates inflationary periods, is given by
\begin{equation}\label{EQN:HSR}
\epsilon\equiv-\frac{\dot{H}}{H^2}= \frac{3}{2}(1+x-y) = 3 x,
\end{equation}
where I have used the constraint, Eq.~(\ref{EQN:Constraint}), in the final equality in Eq.~(\ref{EQN:HSR}). Thus, since $\epsilon < 1$ if and only if $\ddot{a}>0$ (which is the condition for inflation) we have that
\begin{equation}\label{EQN:InflationEpCond}
\textrm{{\it inflation occurs} if and only if }x < 1/3. 
\end{equation}
Thus $x$ is the dynamical variable that will track the occurrence of inflation.

We can provide a useful interpretation for the variables displayed in Eq.~(\ref{EQN:IF}) in the case where the underlying inflationary model corresponds to an SSF model. Invoking the correspondences in Eq.~(\ref{EQN:VLphidotc}), and comparing these with the first Einstein equation [Eq.~(\ref{EQN:EE1})], we see that $x$ corresponds to the fraction of the total energy density that is in the form of kinetic degrees of freedom, whereas $y$ corresponds to the fraction of the total energy density that is in the form of potential degrees of freedom. The variable $\lambda_{m}$ is the fractional change in the {\it m}th derivative of the potential-energy density per unit Hubble time. Salient aspects of inflationary dynamics can indeed be described through an interplay between kinetic and potential degrees of freedom, so these variables comprise a useful choice for an EFT dynamical system that aims to analyze inflation.

Equations of motion for these dynamical variables reveal an infinite tower of first-order, coupled, nonlinear ordinary differential equations---one differential equation for each variable [listed in Eq.~(\ref{EQN:IF})]. Of course, one must close the system of equations to obtain a well-defined, finite-dimensional phase space, and there is a straightforward way to do this. For some $M\geq 0$, one can fix $\lambda_{M}$ to be a constant, thereby obtaining a closed system of differential equations for the dynamical variables $x,y,\lambda_0, \lambda_1, \dots, \lambda_{M-1}$. [In the case where we choose $M=0$ (thereby fixing $
\lambda_{0}\textrm{ = constant}$), a closed system of differential equations arises for just $x$ and $y$.] The resulting phase space is thus $2+M$-dimensional and the corresponding dynamical system is known as the $M$th-order system. 

One can thus study the underlying physical system---one phase space at a time---where there are phase spaces for each dimensionality greater than or equal to two. [Recall the existence of a constraint, Eq.~(\ref{EQN:Constraint}), so that the effective dimensionality of each phase space is reduced by one.] Higher-dimensional phase spaces display richer dynamical behavior, but are also relatively arbitrary. The arbitrariness arises from the fact that a higher-dimensional phase space is derived by fixing $\lambda_{M}\textrm{ = constant}$, for some large $M$. That is, one needs to precisely specify a quantity related to a high derivative of one of the free functions in the action for the EFT of inflation (namely, $L$), and one's warrant for doing so is unclear. A natural  way to proceed is to thus probe the physical system by focusing on the simplest (that is, lowest-dimension) phase spaces first. 

\begin{figure*}
\begin{minipage}{.49\linewidth}
\centering
\subfloat[]{\includegraphics[scale=0.6]{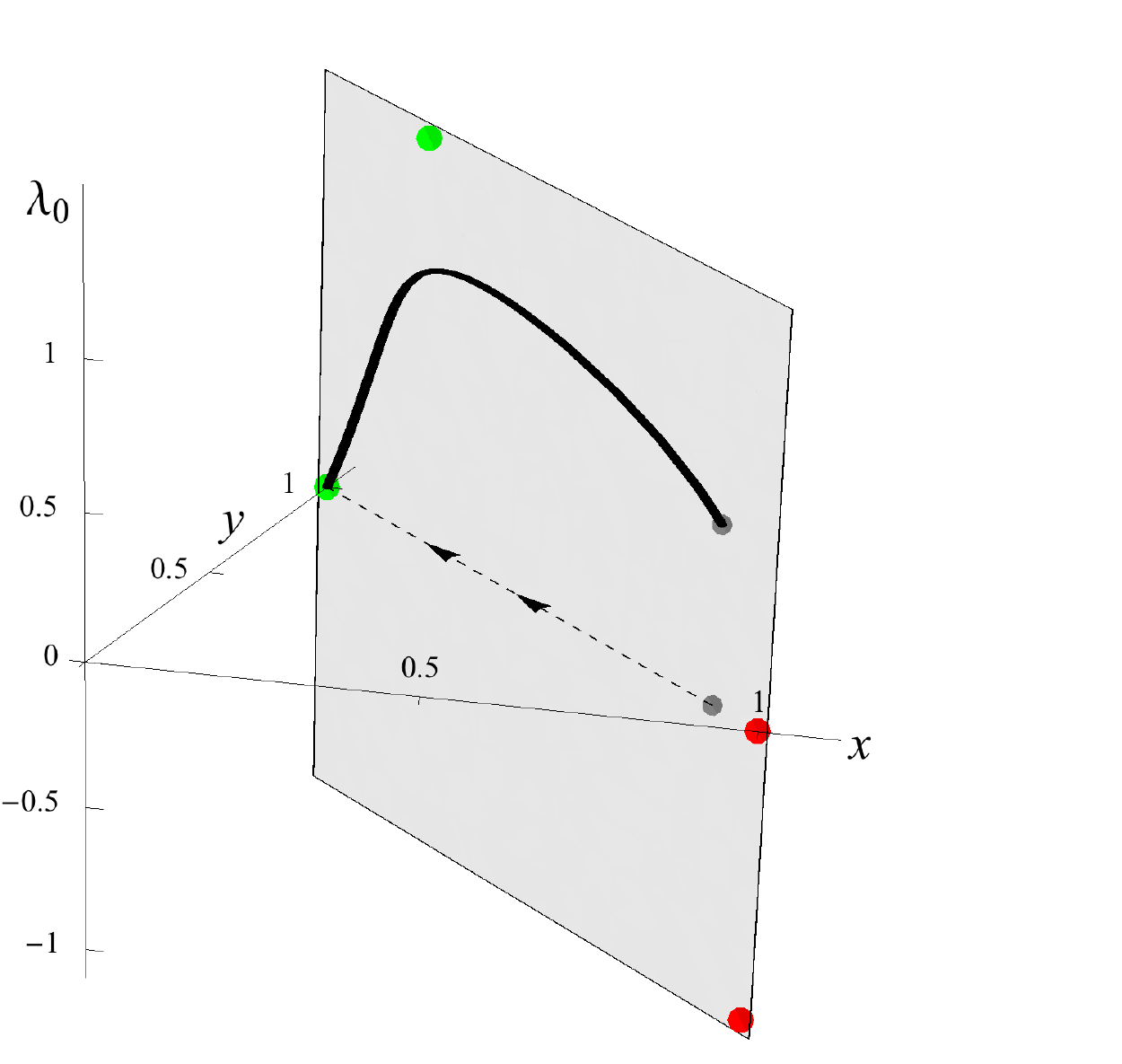}}
\end{minipage}
\begin{minipage}{.49\linewidth}
\centering
\subfloat[]{\includegraphics[scale=0.6]{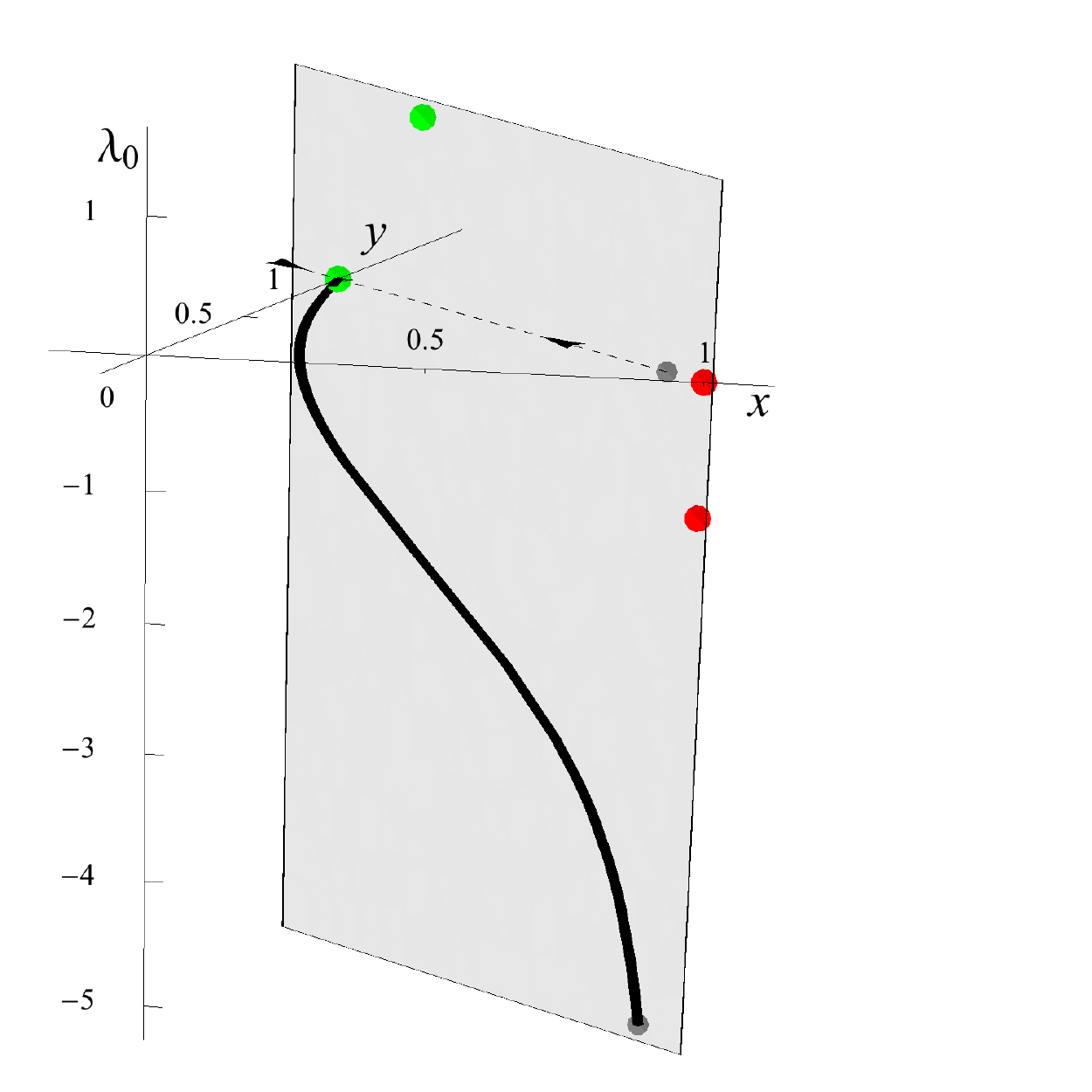}}
\end{minipage}
\begin{minipage}{.49\linewidth}
\centering
\subfloat[]{\includegraphics[scale=0.5]{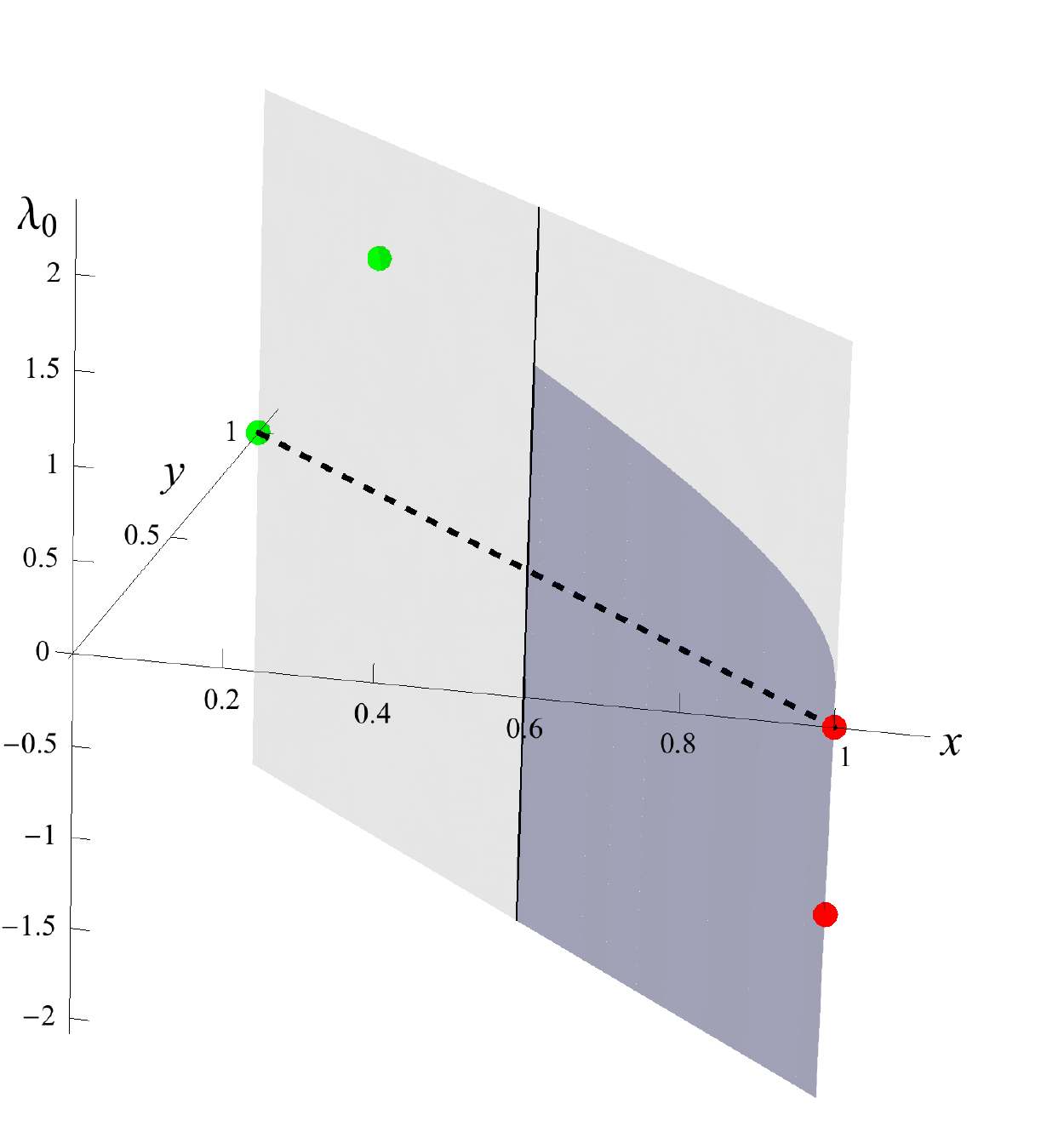}}
\end{minipage}
\begin{minipage}{.49\linewidth}
\centering
\subfloat[]{\includegraphics[scale=0.65]{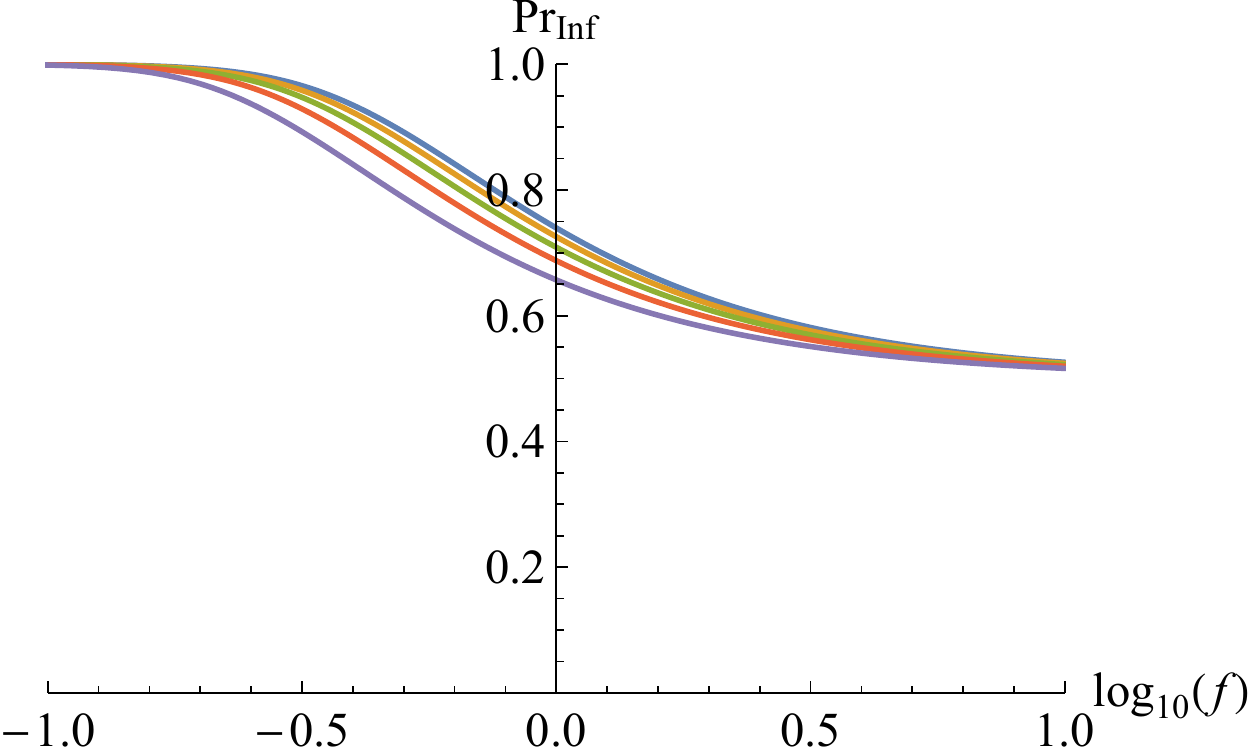}}
\end{minipage}
\caption{An illustrative first-order EFT dynamical system as derived by~\citet{azhar+kaiser_18}. Further details for each subplot are provided in the main text. Subplots (a)--(b) show representative trajectories for background spacetimes that begin with initial conditions that do not favor inflation, which then subsequently flow into inflationary states (where $x<1/3$). The green dots represent inflationary fixed points, whereas the red dots represent noninflationary fixed points. Each trajectory begins at the gray dot and moves along the constraint surface (so that Einstein's equation [Eq.~(\ref{EQN:EE1})] is satisfied), towards an inflationary fixed point. The projection of the trajectory onto the $x$--$y$ plane is shown as a dotted line (with arrows indicating the direction of the flow). The trajectory that starts from a negative initial value of $\lambda_0$ [in (b)] corresponds to a physical setting that lies outside the usual single-scalar-field models of inflation and so reveals novel dynamics---unique to the EFT of inflation formalism. Subplot (c) displays the basin of sufficient inflation, namely, all those initial conditions on the constraint surface that (i) begin dominated by kinetic-energy density and (ii) flow through at least 60 $e$-folds of inflation. Subplot (d) displays the probability of inflation as described in the text and as computed from the basin of sufficient inflation. The five separate curves correspond to five separate choices for the minimum initial value of $x$. The parameter $f$ controls the width of the probability distribution set down over initial conditions, where a larger $f$ corresponds to a broader probability distribution. In each case, there are values of $f$ that lead to a significant probability that sufficient amounts of inflation can occur.
}
\label{FIG:FirstOrderCollage}
\end{figure*}

In Fig.~\ref{FIG:FirstOrderCollage}, I present some key results [from~\citet{azhar+kaiser_18}] as they relate to the initial conditions problem for inflation for a first-order system (which is the next-to-simplest system one can analyze). 
In particular, Figs.~\ref{FIG:FirstOrderCollage}(a) and~\ref{FIG:FirstOrderCollage}(b) show the first-order phase space (that arises by fixing $\lambda_{1}=2$) with two different types of trajectories (in black), distinguished by two different choices of initial conditions. The gray surface corresponds to the constraint surface (that is, where $1=x+y$). Each trajectory starts with (what one might interpret as) kinetic-energy-density-dominated initial conditions, that is, initial conditions that {\it do not favor inflation} (where, initially, $x\geq y$) and then subsequently flows into an inflationary state [which persists as long as $x<1/3$---see Eq.~(\ref{EQN:InflationEpCond})]. Both of the trajectories shown flow towards an inflationary fixed point (the lower green dot, which is an attractor in each case). The purple region shown in Fig.~\ref{FIG:FirstOrderCollage}(c) corresponds to that part of the constraint surface that houses initial conditions that satisfy two criteria: (i) initial conditions are kinetic-energy-density dominated (namely, $x\geq y$) and (ii) trajectories that start at those initial points give rise to at least 60 $e$-folds of inflation (note that the purple region extends to infinity in the negative $\lambda_{0}$-direction). 

In Fig.~\ref{FIG:FirstOrderCollage}(d), I display the probability of flowing through sufficient amounts of inflation (there labeled $\textrm{Pr}_{\textrm{Inf}}$), starting from kinetic-energy-density dominated initial conditions. The computation is largely heuristic (but instructive), in that we compute $\textrm{Pr}_{\textrm{Inf}}$ assuming a particular functional form for the probability distribution over the portion of the phase space in Fig.~\ref{FIG:FirstOrderCollage}(c) that houses kinetic-energy-dominated initial conditions. This probability distribution is uniform along planes parallel to the $x$-$y$ plane, and Gaussian in the $\lambda_0$-direction. This distribution thus disfavors large values of $\lambda_0$, namely, it disfavors large fractional changes in the first derivative of $L$ (per unit Hubble time).  
The five curves in Fig.~\ref{FIG:FirstOrderCollage}(d) correspond to five different choices of the minimum $x$-value over which the probability distribution is defined. This minimum value is chosen to take five different values, namely, $x_{\textrm{min}}=0.5,0.6,0.7,0.8$ or $0.9$, thereby giving rise to five separate curves for the probability of flowing through sufficient amounts of inflation. Each curve is a function of $f$, which is a parameter that determines the width of the Gaussian fall-off along the $\lambda_0$-direction [see~\citet{azhar+kaiser_18} for further details].

We find that the highest probabilities occur for lower values of $x_{\textrm{min}}$. That is, for initial conditions such that the initial kinetic-energy density is less dominant, the probability of flowing through sufficient amounts of inflation is higher. Moreover, for any $x_{\rm min}$, the probability of sufficient inflation increases as $f$ decreases. This is because the width of the probability distribution over initial conditions becomes smaller as $f$ does, in which case a relatively greater amount of the support of the probability distribution comes from initial conditions that lead to trajectories that flow through sufficient amounts of inflation [that is, more of the weight of the probability distribution lies over the purple region in Fig.~\ref{FIG:FirstOrderCollage}(c)].

This condensed description of results represents a small selection of the rather rich set of phase spaces that one can indeed construct from the EFT of inflation described in Sec.~\ref{SEC:EFTI}. The trajectories that one finds display a variety of distinct features, together with surprising universality, when one focuses only on those trajectories that indeed can be mapped onto SSF models of inflation. I leave a more complete account of such technical aspects to a closer inspection of the work by~\citet{azhar+kaiser_18}, save to reiterate the following important conceptual feature. The dynamics as encoded in the EFT of inflation above go well beyond SSF models. Indeed, one can show that trajectories such as those presented in Fig.~\ref{FIG:FirstOrderCollage}(b), which enter into (and remain in) a region where $x<0$ have no analog in the SSF case. In fact, it appears that each initial condition that lies below the $x$--$y$ plane in  Fig.~\ref{FIG:FirstOrderCollage}(c), that is, where the initial value of $\lambda_{0}$ is less than zero, gives rise to a trajectory that does not have an analog in the SSF case. The EFT of inflation combined with the above dynamical-systems analysis encodes general features of single-clock inflationary systems, and thus such a characterization of inflation presents a significantly more comprehensive setting in which to analyze the initial conditions problem for inflation. 

\section{The road ahead}\label{SEC:Road}

There remain a variety of issues that the EFT-of-inflation framework must confront before results in the spirit of those in Sec.~\ref{SEC:InfTraj} can be used to make more definitive claims about the initial conditions problem. In particular, corresponding to (Characterize) and (Measure) in Sec.~\ref{SEC:EFTICP}, there are two shortcomings that EFT-based analyses must overcome---both of which are evident in the analysis presented in Sec.~\ref{SEC:InfTraj}. The first relates to constraints on EFT-based characterizations of inflation that have been developed thus far, in particular, to the degree of homogeneity and isotropy that is built-in to such characterizations. The second is the persistent  problem of extracting a measure, from which probabilistic claims may be generated. 

With respect to the first shortcoming, EFTs of inflation have thus far employed homogeneous and isotropic background spacetimes, about which perturbative expansions are constructed. These expansions incorporate (i) small metric fluctuations (that is, small variations in the geometry of the background spacetime) and (ii) terms that systematically treat higher-order derivatives of the metric---where terms with higher derivatives contribute successively smaller amounts to the action (and thus have less of an influence on the dynamics). The concern here is that if one wishes to probe how generic inflation is, the spacetimes that EFTs of inflation probe do not sample a suitably general set of possible spacetimes. 

Indeed, one of the early prominent criticisms of inflation, as emphasized by~\citet{penrose_89}, was that  a largely homogeneous and isotropic spacetime with small fluctuations is a very special state. Penrose argues that a more generic spacetime would contain much larger inhomogeneities that would presumably collapse into a sea of black holes as the universe evolved in time. 
And, he claims, were inflation able to get started in such a setting, the resulting spacetime that would obtain would not look like the one we observe today. 

Now, the EFT of inflation framework was not expressly designed to probe the initial conditions problem. Rather it was meant as a generic characterization of inflationary dynamics as we believe they occurred. Nevertheless, Penrose's concerns do apply. The spacetimes probed by the EFT of inflation are a small set of possible spacetimes, and so for such model-independent characterizations of inflation to furnish a more exhaustive exploration of the initial conditions problem there remains much work to do. However, part of the promise of EFTs of inflation that I wish to highlight is that (a significant part of) this work may well be tractable. An analog of recent numerical work by, for example,~\citet{east+al_16} and~\citet{clough+al_16, clough+al_18}, but for EFT-based approaches, would be an interesting next step. As would more analytical approaches that construct, for example, EFTs of inflation that include background spacetimes that are anisotropic, for which there is already a useful classification scheme---namely, the Bianchi classification scheme~\citep{bianchi_98}. 

Regardless of the precise construction employed in some more generic EFT of inflation, the measure problem over initial conditions for the underlying cosmological spacetime persists, as introduced in (Measure) in Sec.~\ref{SEC:EFTICP}. The probability distribution we invoked for the first-order system in Sec.~\ref{SEC:InfTraj} was ad hoc. It was designed to provide a measure over initial conditions in such a way as to allow one to examine, in a familiar way (that is, by varying the width of a Gaussian), the dependence of the probability of flowing through sufficient amounts of inflation on the effective size of the region of initial conditions considered. Of course, quantitatively different probability distributions over initial conditions would give different values for this probability.\footnote{In~\citet{azhar+kaiser_18}, we did probe another type of distribution (a boxlike distribution) that implements a hard cut-off in the $\lambda_0$-direction. Such a distribution gives weight to a smaller subset of initial conditions (compared to a distribution with a similar width that is Gaussian in the $\lambda_0$-direction) and, in particular, to a smaller subset of initial conditions that don't lead to inflation. This leads to modestly higher values for the probability of sufficient inflation.} And as stated in Sec.~\ref{SEC:EFTICP}, there is no agreed-upon solution to this measure problem. Perhaps the state of the art is represented by work by~\citet{remmen+carroll_13, remmen+carroll_14}, which looks at issues of measure in the context of SSF models of inflation. Their work builds on earlier attempts to construct measures on spaces of trajectories, such as by~\citet{gibbons+al_87}. The measure of~\citet{gibbons+al_87}, however, diverges for flat FLRW universes [as indeed noted by:~\citet{hawking+page_88, coule_95, gibbons+turok_08, carroll+tam_10, remmen+carroll_13, remmen+carroll_14, carroll_14}], and the work of~\citet{remmen+carroll_13, remmen+carroll_14} finds a way around this issue by explicitly constructing measures over trajectories for flat FLRW universes in the context of specific SSF potentials. Yet there remains much work to do in extending such measures to more generic background spacetimes that involve SSF models of inflation (let alone to model-independent EFT-based characterizations of inflation).

So a definitive statement about how generic inflation is, would be difficult to obtain even if we could write down a model-independent EFT-based description of inflation that is more general than those that have been currently developed. Of course, an important consideration is the role such a  statement would have in our assessment of the theory of inflation. This issue touches upon a topic that I do not have space to enter into here, related to norms for the assessment of theories in the sciences more generally. In particular, the key question is whether a lack of fine-tuning is an important element in the assessment and development of theories in physics. [See~\citet{azhar+loeb_18, azhar+loeb_19} for a treatment of such an issue for theories (and models) considered broadly across the sciences, wherein we argue that a lack of fine-tuning {\it is} an important consideration.]

And so, with a more operational approach in mind (for now), there are promising avenues for further work that will help in determining how generic cosmic inflation may be. One such avenue is to construct an expressly Hamiltonian formulation of an EFT of inflation. Such a Hamiltonian formulation is, indeed, not the formulation described in Sec.~\ref{SEC:InfTraj}, but does underlie the work by~\citet{gibbons+al_87} and~\citet{remmen+carroll_13, remmen+carroll_14} (for FLRW cosmologies). The goal would be to construct, then, a conserved measure over the space of trajectories of an EFT of inflation that incorporates {\it significant} anisotropies and inhomogeneities. Of course, infinite measures would need to be regulated in some way, that is, at the very least some {\it conventional} solution to the measure problem would need to be established. Were such a construction possible: (i) it would complement recent work that has been carried out on the initial conditions problem, as described above [see (Characterize) in Sec.~\ref{SEC:EFTICP}, and Sec.~\ref{SEC:InfTraj}]; but (ii) by virtue of the model-independent approach inherent in EFTs of inflation, it would provide an important perspective on the question of the degree to which the theory of inflation is finely tuned. 

\section{Envoi}\label{SEC:Conclusion}

I will conclude by emphasizing my central claim, starting with a brief summary of some of the themes that underlie this claim. 

An important guide to our assessment of a theory in the sciences more generally is the degree of fine-tuning it exhibits. How we characterize a theory and how we conceptualize and compute levels of fine-tuning are key elements of such an assessment. For cosmic inflation, recently developed EFTs of inflation provide a thoroughly novel way to think about the theory. In particular, inflation is characterized via the symmetries of the underlying spacetime that result from considerations of fundamental physical facts about inflation: that inflation is a period of accelerated expansion of the (observable) universe, where small density fluctuations arise, and where inflation indeed ends at some point in time. Thus inflation is characterized as a theory in which time-translation invariance is broken, and where there is a single clock that, in effect, counts down time until the end of inflation. Crucially, the nature and physical underpinnings of this single clock do not have to be specified. This allows one to construct an action for the theory in very general terms, so as to include all SSF models of inflation. Then, as described above [and especially in~\citet{azhar+kaiser_18}], there are ways in which one can specify initial conditions for EFTs of inflation, so that one can analyze the flow into inflationary states of such a model-independent characterization of inflation. 

An obstacle to addressing the initial conditions problem for inflation (as introduced in Sec.~\ref{SEC:ICP}) is that characterizing fine-tuning---in particular, finding a quantitative account of fine-tuning---is a difficult task (and this feature of the problem extends to characterizing fine-tuning across the sciences more broadly). Thinking of fine-tuning in measure-theoretic terms one generally runs into a measure problem, where a unique measure or probability distribution over the initial conditions of the theory is not singled out (regardless of how one characterizes inflation). However there may be measures that are at least well-motivated, rudiments of which have been studied in the context of SSF models of inflation, which could be adapted for use in the context of EFTs of inflation. 

If such a measure problem has (at least) a conventional solution, then the central claim in this paper would be established. For the combination of a new way of thinking about inflation (via general EFTs of inflation---more general than those that have been constructed thus far), with the adaptation of familiar (measure-theoretic) tools to this new context, would provide a thoroughly original and relatively definitive way to characterize fine-tuning of cosmic inflation.  

\addcontentsline{toc}{section}{Acknowledgements}
\section*{Acknowledgements}
I am very grateful to John Barrow, Jeremy Butterfield, Peter Galison, Alan Guth, Peter Koellner, Andrei Linde, Abraham Loeb, Mohammad Hossein Namjoo, Thomas Ryckman, Chris Smeenk, and Nicholas Teh, for helpful discussions in various settings and at various stages of the work presented herein. I wish to especially thank David Kaiser for illuminating discussions and joint work related to this paper. I acknowledge support from the Black Hole Initiative at Harvard University, which is funded through a grant from the John Templeton Foundation and the Gordon and Betty Moore Foundation. The views presented in this paper do not necessarily reflect those of any person or funding agency mentioned above.

\end{document}